\documentclass[twocolumn,showpacs,preprintnumbers,amsmath,amssymb]{revtex4}

\input psfig.sty

\usepackage{graphicx}
\usepackage{dcolumn}
\usepackage{bm}

\begin{document}


\title{Deflationary cosmology: constraints from angular size and ages of globular clusters}

\author{J. V. Cunha$^{1}$} \email{jvital@dfte.ufrn.br} 

\author{J. S. Alcaniz$^{2}$} \email{alcaniz@astro.washington.edu}

\author{J. A. S. Lima$^{1}$} \email{limajas@dfte.ufrn.br}
\affiliation{
$^{1}$Departamento de F\'{\i}sica, Universidade Federal do Rio Grande do
Norte, C.P. 1641, 59072-970, Natal, RN, Brasil \\
$^{2}$Astronomy Department, University of Washington,
Seattle, Washington, 98195-1580, USA}

\date{\today}

\begin{abstract}
Observational constraints to a large class of decaying vacuum
cosmologies are derived using the angular size data of compact
radio-sources and the latest age estimates of globular clusters.
For this class of deflationary $\Lambda(t)$ models, the present
value of the vacuum energy density is quantified by a positive
$\beta$ parameter smaller than unity. In the case of
milliarcsecond compact radio-sources, we find that the allowed
intervals for $\beta$ and the matter density parameter $\Omega_m$
are heavily dependent on the value of the mean projected linear
size $l$. For $l \simeq 20h^{-1} - 30h^{-1}$ pc, the best fit
occurs for $\beta \sim 0.58$, $\Omega_{\rm{m}} \sim 0.58$, and
$\beta \sim 0.76$, $\Omega_{\rm{m}} \sim 0.28$, respectively. This
analysis shows that if one minimizes $\chi^{2}$ for the free
parameters $l$, $\Omega_{\rm{m}}$ and $\beta$, the best fit for
these angular size data corresponds to a decaying $\Lambda(t)$
with $\Omega_{\rm{m}} = 0.54$ $\beta=0.6$ and $l = 22.64h^{-1}$
pc. Constraints from age estimates of globular clusters and old
high redshift galaxies are not so restrictive, thereby suggesting
that there is no age crisis for this kind of $\Lambda(t)$
cosmologies.
\end{abstract}

\maketitle

\section{Introduction}

The search for alternative cosmologies is presently in vogue. The
leitmotiv is the observational support for an accelerated Universe
provided by the type Ia supernovae (SNe) experiments at
intermediate and high redshifts \cite{perlmutter,riess}. Such
observations are consistently indicating that the bulk of energy
in the Universe is repulsive and appears like a quintessence
component, that is, an unknown form of energy (in addition to the
ordinary dark matter), probably of primordial origin \cite{sahni}. In the current literature there 
are many candidates for this negative-pressure dark component, among them: (i) a vacuum decaying 
energy density, or
a time varying $\Lambda(t)$ \cite{ozer}, (ii) a time varying relic
scalar field slowly rolling down its potential \cite{ratra},
(iii) the so-called ``X-matter", an extra component simply
characterized by an equation of state $p_x=\omega\rho_{x}$, where
$\omega\geq -1$ \cite{turner} and (iv) models based on the framework of brane-induced gravity
\cite{dvali}.

In a recent paper \cite{joao} (henceforth paper I), we derived the
basic expressions for kinematic tests in a quintessence scenario
of the type (i) above, which has originally been termed
deflationary cosmology \cite{jackson,lima1}. The effective time
dependent cosmological term is regarded as a second fluid
component with an energy density, $\rho_v(t) =\Lambda(t) /{8 \pi G}$,
which transfers energy continuously to the material component.
Such a scenario has an interesting cosmological history evolving
in three stages. First, an unstable de Sitter configuration is
supported by the largest value of the decaying vacuum energy
density. This nonsingular state evolves to a quasi-Friedmann-Roberton-Walker (FRW) 
vacuum-radiation phase and, subsequently, the Universe approaches
continuously the present vacuum-dust stage. The first stage
harmonizes the scenario with the cosmological constant problem,
while the transition to the second stage solves the horizon and
other well-know problems in the same manner as in inflation.
Finally, the Universe enters in the present accelerated
vacuum-dust phase required by the SN Ia observations.

More recently, it has been argued that deflation can be described
in terms of a scalar field coupled to the fluid component
\cite{jackson1}. Such models are analytic examples of warm
inflationary scenarios proposed by Berera \cite{WI} where particle
production occurs during the evolution. As a consequence, the
supercooling process, as well as the subsequent reheating are no
more necessary. In this concern, it should be interesting to
analyze the generation of the perturbation spectra by comparing
the results with the ones recently obtained by Taylor and Berera
for warm inflation \cite{TB00}. Here, as there, one may expect a
suppression of the tensor-to-scalar ratio perturbations, though a
detailed study is necessary in order to have a more definitive
conclusion. On the other hand, the scalar field in this enlarged framework is
also thermally coupled with dark matter, and as such, it can be
used to avoid the cosmic coincidence problem. Specific examples of
exact deflationary models and the underlying thermodynamics has
been given by Gunzig {\it et al.} \cite{GMN}, and some scalar
field motivated descriptions were also investigated in detail by
Maartens {\em et al.} \cite{Maartensetal} and Zimdahl
\cite{Zimdahl}.

In this paper we discuss more quantitatively how the
observations put limits on the free parameters of the deflationary
scenario. We focus our attention on constraints from the angular size
of milliarcsecond compact radio sources, and the latest age
estimates of globular clusters (GCs), and for completeness we also
consider shortly the case for old high redshift galaxies (OHRGs).
Next section we set up the basic equations for these models. In
Section III we derive the corresponding constraints for deflationary
cosmologies from $\theta(z)$ analysis. Limits from GCs and OHRGs
are discussed in Section IV. The main results are summarized in the
conclusion section.

\section{Basic equations}

Let us now consider a class of spacetimes described by the general
FRW line element ($c=1$)
\begin{equation}
ds^2 = dt^2 - R^{2}(t) \left[d\chi^{2} + S^{2}_{k}(\chi) (d
 \theta^2 + \rm{sin}^{2} \theta d \phi^{2})\right]  ,
\end{equation}
where $\chi$, $\theta$, and $\phi$ are dimensionless comoving
coordinates, $R(t)$ is the scale factor, and $S_{k}(\chi)$ depends
on the curvature parameter ($k=0$, $\pm 1$). The later function is
defined by one of the following forms: $S_k (\chi) = \rm{sinh}
(\chi)$, $\chi$, $\rm{sin} \chi$, respectively, for open, flat and
closed universes. Since the main interest here is related to
kinematic tests for deflation, in what follows we consider only
the Einstein field equations (EFE) for a nonvacuum pressureless
component plus a cosmological $\Lambda(t)$-term:

\begin{equation}
\label{rho} 8\pi G \rho + \Lambda(t) = 3 \frac{\dot{R}^2}{R^2} + 3
\frac{k}{R^2} \quad,
\end{equation}
\begin{equation} \Lambda(t) = -2 \frac{\ddot{R}}{R} -
\frac{\dot{R}^2}{R^2} - \frac{k}{R^2}\quad ,
\end{equation}
where an overdot means time derivative and $\rho$ is the dust
energy density. The effective $\Lambda (t)$ is a variable dynamic
degree of freedom. In our context, it relaxes to the present
value, $\Lambda_o$, according to the deflationary ansatz
\cite{lima1}
\begin{equation} \label{ansatz}
\rho_{v} = \frac{\Lambda(t)}{8 \pi G} = \beta \rho_{T} \left(1 +
\frac{1 - \beta}{\beta} {H \over H_{I}}\right) \quad ,
\end{equation}
where $\rho_{v}$ is the vacuum density, $\rho_{T}=\rho_{v}+ \rho$
is the total energy density, $H={\dot{R}}/R$ is the Hubble
parameter, $H_{I}^{-1}$ is the arbitrary time scale characterizing
the deflationary period, and $\beta\in[0,1]$ is a dimensionless
parameter of order unity.

Deflationary universe models start their evolution with the
largest value of the Hubble parameter, $H=H_{I}$, for which the
phenomenological expression (4) reduces to $\rho_{v}=\rho_{T}$ so
that we have inflation with no matter and radiation components
($\rho=0$). At late times ($H << H_{I}$), we see that $\rho_{v}
\sim \beta \rho_{T}$, as required by the recent observations.
Therefore, if the deflationary process begins at Planck time one
has $H_{I}^{-1}\sim 10^{-43}$s, and since $H_{0}^{-1}\sim
10^{17}$s it thus follows that $H_{0}/H_{I} \sim 10^{-64}$ while
the remaining terms are of order unity. Such a conclusion does not
change appreciably if the Planck scale is replaced by the grand or
even the electroweak unification scales in accordance to the
standard model (see paper I). This means that to a high degree of
accuracy, the scale $H_{I}$ is unimportant during the vacuum-dust
dominated phase so that for all practical purposes the vacuum
energy density can be approximated by $\rho_v = \beta \rho_{T}$.

An interesting feature of decaying $\Lambda(t)$  models is that
the temperature dependence on the redshift $z$ of the relic
radiation, $T(z)$, can be different from the standard prediction,
which is deduced assuming an adiabatic expansion. For deflationary
universes, the temperature law at the vacuum/dust phase reads
\begin{equation}
T(z) = T_o(1 + z)^{1 - \beta}.
\end{equation}
where $T_o$ is the temperature of the Cosmic Microwave Background (CMB) at $z = 0$. This new
relation implies that for a given redshift $z$, the temperature of
the Universe is slightly lower than in the standard
photon-conserved scenario. Actually, indirect measurements of
$T(z)$ at high redshifts \cite{S94,GE97,los} may become one of the
most powerful cosmological tests because it may exclude the
presence of a cosmological constant or even of any kind of separately
conserved quintessence \cite{LAV00}.

\begin{figure}
\vspace{.2in}
\centerline{\psfig{figure=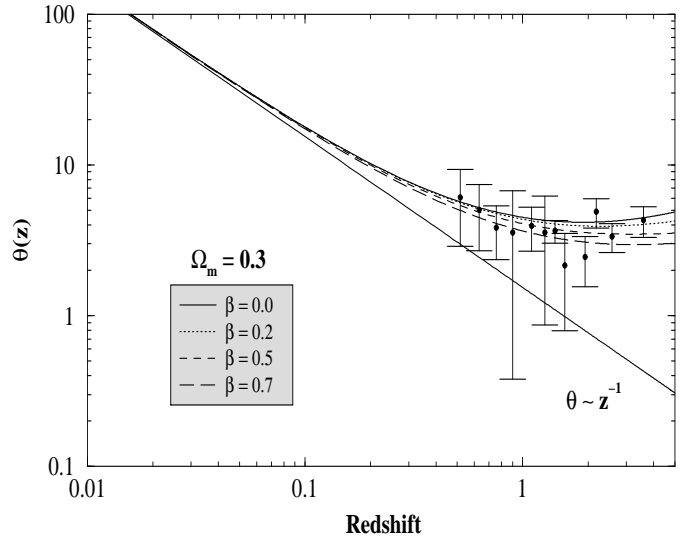,width=3.5truein,height=2.9truein}
\hskip 0.1in} \caption{The angular size - redshift relation for
145 sources binned into 12 bins [25]. The curves correspond to
$\Omega_{\rm{m}} = 0.3$ and $l = 26.46 h^{-1}$ pc.}
\end{figure}

\section{Constraints from angular size-redshift diagram}

The angular size - redshift relation to a light source of proper
size $l$ (assumed free of evolutionary effects) can be obtained by
integrating the spatial part of Eq. (1) for $\xi$ and $\phi$
fixed. One finds
\begin{equation}
\theta = \frac{D(1 + z)}{\xi(z)}\quad.
\end{equation}
In the above expression $D$ is the angular-size scale expressed in
milliarcseconds (mas)
\begin{equation}
D = 100 lh ,
\end{equation}
where $l$ is measured in parsecs (for compact radio-sources). The
dimensionless coordinate $\xi$ is given by (see Appendix A of
paper I)
\begin{equation} \label{q2}
\xi(z) = \int_{(1+z)^{-1}}^{1} {dx \over x\left[
1-(\frac{\Omega_{\rm{m}}}{1-\beta}) +
(\frac{\Omega_{\rm{m}}}{1-\beta}) x^{-(1-3\beta)} \right]^{1/2}}.
\end{equation}
Now, by integrating the previous expression and inserting the
result into Eq.(6) one finds
\begin{equation}
\theta = \frac{D \left(\frac{\Omega_{\rm{m}}}{1 -
\beta}\right)^{1/2} (1 + z)}{{{\rm{sin}}\left[\delta
sin^{-1}\alpha_1-\delta sin^{-1}\alpha_2\right]}}\quad .
\end{equation}
where $\delta=\frac{2}{(1-3\beta)}$, $\alpha_1=
(1-\frac{1-\beta}{\Omega_{\rm{m}}})^{{1 \over 2}}$, and
$\alpha_{2}=\alpha_{1} (1 + z)^{-(\frac{1-3\beta}{2})}$.

The above equations show that the predicted value of $\theta(z)$
is completely determined once the values of $l$, $\Omega_{\rm{m}}$
and $\beta$ are given. Two points, however, should be stressed
before discussing the resulting diagrams. First of all, the
predicted values of $\Omega_{\rm{m}}$ and $\beta$ are strongly
dependent on the adopted value of $l$. In the absence of a
statistical study describing the intrinsic length distribution of
the sources, we consider here the approach recently discussed by
Lima and Alcaniz \cite{lima}. More specifically, instead of
assuming a given value for the mean projected linear size, we work
on the interval $l \simeq 20h^{-1} - 30h^{-1}$ pc, i.e., $l \sim
O(40)$ pc for $h = 0.65$, or equivalently, $D = 1.4 - 2.0$ mas. In
addition, following Kellermann \cite{kell}, we also assume that
compact radio sources are free of the evolutionary and selection
effects that have bedevilled attempts to use extended double radio
source in this context (see, for example, \cite{bucha}), as they
are deeply embedded in active galactic nuclei, and, therefore,
their morphology and kinematics do not depend considerably on the
changes of the intergalactic medium. Moreover, these sources have
typical ages of some tens of years, i.e., it is reasonable to
suppose that a stable population is established, characterized by
parameters that do not change with the cosmic epoch \cite{jack}.

\begin{figure}
\vspace{.2in}
\centerline{\psfig{figure=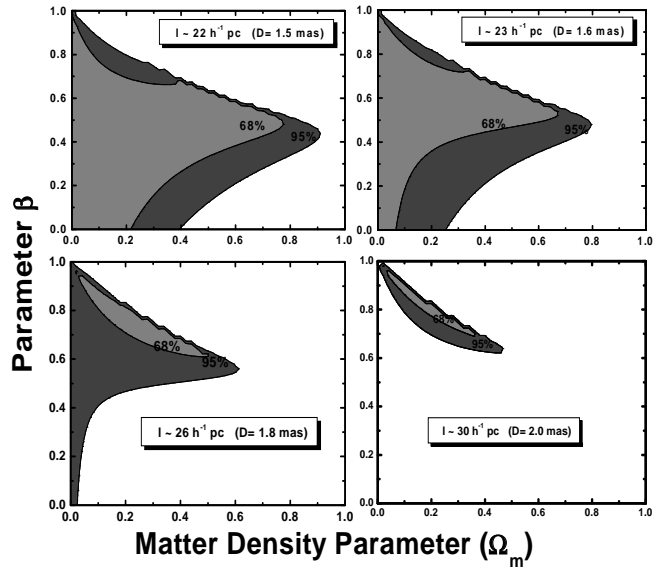,width=3.5truein,height=3.0truein}
\hskip 0.1in} \caption{Confidence regions in the
$\beta$-$\Omega_{\rm{m}}$ plane according to the updated sample of
angular size data [25]. The solid lines show the $95$\% and $68$\%
likelihood contours for decaying vacuum models.}
\end{figure}

In our analysis we consider the angular size data for
milliarcsecond radio-sources recently compiled by Gurvits {\it et
al.} \cite{gurv}. This data set, originally composed by 330
sources distributed over a wide range of redshifts ($0.011 \leq z
\leq 4.72$), was reduced to 145 sources with spectral index $-0.38
\leq \alpha \leq 0.18$ and total luminosity $Lh^{2} \geq 10^{26}$
W/Hz in order to minimize any possible dependence of angular size
on spectral index and/or linear size on luminosity. This new
sample was distributed into 12 bins with 12-13 sources per bin. In
order to determine the cosmological parameters $\Omega_{\rm{m}}$
and $\beta$, we use a $\chi^{2}$ minimization for a range of
$\Omega_{\rm{m}}$ and $\beta$ spanning the interval [0,1] in steps
of 0.02
\begin{equation}
\chi^{2}(l, \Omega_{\rm{m}}, \beta) =
\sum_{i=1}^{12}{\frac{\left[\theta(z_{i}, l, \Omega_{\rm{m}},
\beta) - \theta_{oi}\right]^{2}}{\sigma_{i}^{2}}},
\end{equation}
where $\theta(z_{i}, l, \Omega_{\rm{m}}, \beta)$ is given by Eq.
(9) and $\theta_{oi}$ is the observed values of the angular size
with errors $\sigma_{i}$ of the $i$th bin in the sample.

\begin{table}
\begin{center}
\begin{tabular}{rrrlll}
\hline \hline \\ \multicolumn{1}{c}{$D$ (mas)}&
\multicolumn{1}{c}{$lh$ (pc)}&
\multicolumn{1}{c}{$\Omega_{\rm{m}}$}&
\multicolumn{1}{c}{$\beta$}& \multicolumn{1}{c}{$\chi^{2}$}\\
\\  \hline  \hline
1.5& 22.05& 0.58& 0.58& 4.30\\
\\
1.6& 23.53& 0.5& 0.62& 4.32\\
\\
1.8& 26.47& 0.34& 0.72& 4.65\\
\\
2.0& 29.41& 0.28& 0.76& 6.05\\
\\
{\bf Best fit:} 1.54& 22.64& 0.54& 0.6& 4.28\\ \hline \hline
\\
\end{tabular}
\begin{center}
\caption{Angular size constraints on varying $\Lambda$
cosmologies}
\end{center}
\end{center}
\end{table}

In Fig. 1 we display the binned data of the median angular size
plotted against redshift for $\Omega_{\rm{m}} = 0.3$ and several
values of $\beta$. The curve for a static Euclidean universe is
also shown for comparison ($\theta \sim z^{-1}$). As can be seen
from the figure, Euclidean models are strongly deprived by the
observational data. In our statistical analysis presented below,
it was implicitly assumed that possible selection effects
(including the resolution angular limit) are completely under
control.

Figure 2 shows contours of constant likelihood (95$\%$ and 68$\%$) in the plane $\Omega_{\rm{m}} -
\beta$ for the
interval $l \simeq 20 - 30h^{-1}$ pc. Note that the allowed range
for $\beta$ is reasonably large unless the characteristic length
is also large ($l \sim 30h^{-1}$ pc). For $l \simeq 22h^{-1}$ pc, the
best fit occurs for a closed model with $\Omega_{\rm{m}} = 0.58$
and $\beta = 0.58$. In the subsequent panels of the same figure
similar analyzes are displayed for $l \simeq 23.53h^{-1}$ pc (D =
1.6 mas), $l \simeq 26.47h^{-1}$ pc (D = 1.8 mas) and $l \simeq
29.41h^{-1}$ pc (D = 2.0 mas). As physically expected, the limits
are now more restrictive for the matter density contribution while
larger values of $\beta$ are allowed. It happens because since
$\theta \sim l/\xi$, for the same data ($\theta_{oi}$) and larger
$l$ we need larger $\xi(z)$ and, therefore, smaller (larger)
values of $\Omega_{\rm{m}}$ ($\beta$). In particular, if we
minimize $\chi^{2}$ for $l$, $\Omega_{\rm{m}}$, and $\beta$, we
obtain $l = 22.64^{-1}$ pc (D = 1.54 mas), $\Omega_{\rm{m}} =
0.54$, and $\beta = 0.6$ with $\chi^{2} = 4.28$ and 9 degrees of
freedom. Such values are also in agreement to the ones recently
found by Vishwakarma \cite{vish}. Indeed, as explained in Paper I,
a subclass of the models investigated by him ($\Lambda = n\Omega
H^{2}$) corresponds to the late stages of our deflationary
scenario. We also remark that although not discussed here, it is
possible to determine exactly the influence of the
$\Lambda(t)$-term in the critical redshift $z_m$ at which the
angular size takes its minimal value. However, as shown elsewhere
\cite{alcaniz}, the critical redshift cannot discriminate between
world models since different scenarios may provide the same $z_m$
values. The main results of this section are summarized in Table
1.

\begin{figure}
\vspace{.2in}
\centerline{\psfig{figure=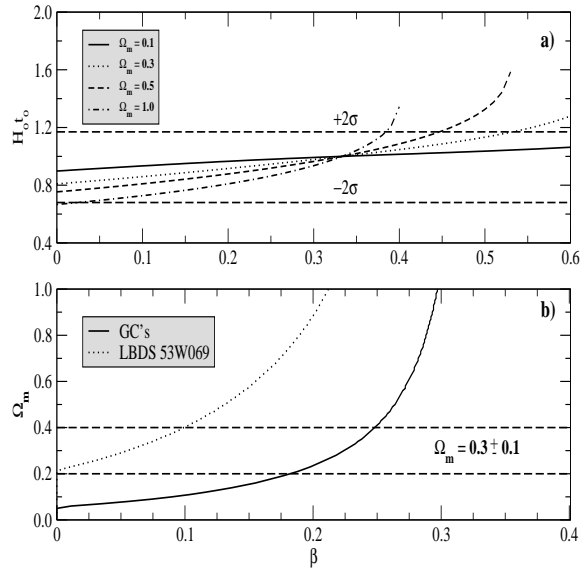,width=3.5truein,height=3.8truein,angle=-90}
\hskip 0.1in} \caption{{\bf a)}$H_ot_o$ as a function of $\beta$
for some values of the density parameter $\Omega_{\rm{m}}$.
Horizontal lines indicate the $\pm2\sigma$ limits of the age
parameter $H_ot_o = 0.93 \pm 0.12$. {\bf b)} Contours of fixed age
parameters. Solid curve corresponds to the median value of the
most recent estimates of globular clusters ($t_o = 13\pm 1$ Gyr)
and dashed line to the radio galaxy LBDS 53W069 which has an
estimated age of $t_z = 4.0$ Gyr at $z = 1.43$.}
\end{figure}

\section{Constraints from ages of globular clusters}

Let us now discuss the constraints on deflationary cosmologies
related to the expanding age of the Universe. As widely known, a
lower bound for this quantity can be estimated in a variety of
different ways. For instance, Oswalt {\it et al.} \cite{osw},
analyzing the cooling sequence of white dwarf stars found a lower
age limit for the galactic disk of 9.5 Gyr. Later on, a value of
15.2 $\pm$ 3.7 Gyr was also determined using radioactive dating of
thorium and europium abundances in stars \cite{cowan}. In this
connection, the recent age estimate of an extremely metal-poor
star in the halo of our Galaxy (based on the detection of the
385.957 nm line of singly ionized $^{238}$U) indicated an age of
12.5 $\pm$ 3 Gyr \cite{cay}.

Probably, the most important way of estimating a lower limit to
the age of the Universe is dating the oldest stars in GCs. Such
estimates have, however, oscillated considerably since the
publication of the statistical parallax measures done by
Hipparcos. Initially, some studies implied in a lower limit of 9.5
Gyr at $95\%$ confidence level (c.l.) \cite{chab}, making some authors argue
immediately for the end of the age problem \cite{krauss}.
Nevertheless, subsequent studies \cite{chab1} using new
statistical parallax measures and updating some stellar model
parameters, found 13.2 Gyr with a lower limit of 11 Gyr at 95$\%$
c.l., as a corrected mean value for age estimates of GCs (see also
\cite{carreta}. This implies that the flat cold dark matter (CDM) model is ruled out
for $h \geq 0.50$, thereby showing that the ``age crisis" is not
dead at all since the most recent measurements of $h$ point
consistently to $h \geq 0.65$ \cite{gio,f1,f2}. All these results
are also in accordance with recent estimates based on rather
different methods for which the ages of the oldest GCs in our
Galaxy fall on the interval 13.8 - 16.3 Gyr \cite{rengel}.

The general age-redshift relation for a deflationary universe is
(see Paper I)
\begin{equation}
t_o = H_o^{-1} \int^{1}_{0} {dx \over \left[1 - {\Omega_{\rm{m}}
\over (1 - \beta)} + {\Omega_{\rm{m}} \over (1 - \beta)} x^{3\beta
- 1}\right]^{1\over 2}}.
\end{equation}
As one may conclude, by fixing the $H_ot_o$ from observations, the
limits on the $\beta$ parameter may readily be obtained for some
specified values of $\Omega_{\rm{m}}$. Note also that the age
parameter depends only on the product of the two quantities $H_o$
and $t_o$, which are measured from completely different methods.
Here we consider $t_o = 13\pm 1$ Gyr as a median value for the
most recent estimates of globular clusters, and, following
Freedman \cite{f2}, we assume $H_o = 70\pm 0.7 \rm{Km.s^{-1}.
Mpc^{-1}}$. Since these quantities ($H_o$ and $t_o$) are
independent, the two errors may be added in quadratures providing
$H_ot_o = 0.93 \pm 0.12$, a value very close to some
determinations based on SN Ia data \cite{riess,tonry}. From this
analysis we see that the Einstein-de Sitter case is off by $\sim$
2 standard deviations.

In Fig. 3a we show the dimensionless product $H_ot_o$ as a
function of $\beta$ for some values of the density parameter
$\Omega_{\rm{m}}$. Horizontal dashed lines indicate $\pm 2\sigma$
of the age parameter for the values of $H_o$ and $t_o$ considered
here. As should be physically expected, the greater the
contribution of the vacuum ($\beta$) the larger the age predicted
by the model. For $\beta = 1/3$, all models predict $H_ot_o = 1$
regardless of the value of $\Omega_{\rm{m}}$. As explained in
Paper I, this critical case corresponds to an expanding universe
with constant Hubble flow ($q_o = 0$) also termed ``coasting
cosmology" \cite{kolb}.

In Fig. 3b we show the plane $\Omega_{\rm{m}} - \beta$ for the
fixed value of the product $H_ot_o = 0.93 \pm 0.12$ (solid curve).
Horizontal lines correspond to the observed range $\Omega_{\rm{m}}
= 0.2 - 0.4$ \cite{calb}, which is used to fix the limits on the
$\beta$ parameter. For this interval we find $\beta \geq 0.18$ and
$\beta \geq 0.25$, respectively. For comparison, we also have
plotted the curve for an object (LBDS 53W069) with 4.0 Gyr at $z =
1.43$ \cite{dunlop}. The estimated age of this radio galaxy does
not provide restrictive limits on the $\beta$ parameter, thereby
showing that deflationary cosmologies are quite efficient to solve
the variant of the age crisis at high-$z$. These results are also
in agreement with those found in Ref. \cite{alc1999} in the
context of the standard and $\Lambda$CDM models.

\section{Conclusion}

We have investigated the observational constraints on deflationary
cosmologies provided by the angular size data and age estimates of
globular clusters. By considering a sample of 145 milliarcsecond
radio-sources recently compiled by Gurvits {\it et al.}
\cite{gurv} we found that the best fit occurs for a closed model
with $\Omega_{\rm{m}} = 0.54$, $\beta = 0.6$ ($\Omega_{\rm{T}} =
1.3$), and a characteristic length of $l = 22.64^{-1}$ pc (D =
1.54 mas). As we have seen, angular size measurements from compact
radio sources may provide an important test for world models
driven by a dark energy component. We stress, however, that
constraints from the angular size redshift relation should be taken
with some caution because a statistical analysis describing the
intrinsic length distribution of the sources is still lacking.
This aspect should be specially investigated since the recent
development of the observational techniques certainly will provide
more accurate data of angular size in the near future.

Concerning the age constraints, by adopting $t_o = 13 \pm 1$ Gyr
as a median value for the most recent age estimates of globular
clusters and $h = 0.7\pm 0.07$ we have showed that deflationary
models are very efficient to solve the ``already" classical age of
the Universe problem, as well as its variant related to old
galaxies at high redshifts. In particular, regardless of the value
of $\Omega_{\rm{m}}$, for $\beta \geq 0.25$ the class of models
studied here provides total ages greater than 14 Gyr.

\begin{acknowledgments}
The authors are grateful to L. I. Gurvits for sending his
compilation of angular size data as well as for helpful
conversations. We also thank the anonymous referee for useful suggestions and discussions. This
work was partially supported by the Conselho
Nacional de Desenvolvimento Cient\'{\i}fico e Tecnol\'{o}gico
(CNPq), Pronex/FINEP (No. 41.96.0908.00), FAPESP (00/06695-0) and CNPq (62.0053/01-1-PADCT 
III/Milenio).

\end{acknowledgments}


\end{document}